# PREDICTING THE FRACTURE PROPENSITY OF AMORPHOUS SILICA USING MOLECULAR DYNAMICS SIMULATIONS AND MACHINE LEARNING

JIAHAO LIU

*Sibley School of Mechanical and Aerospace Engineering,*
*Cornell University, Ithaca, New York, USA*
*Department of Engineering Mechanics,*
*Xi'an Jiaotong University, Xi'an, Shaanxi, China*

JINGJIE YEO

*Sibley School of Mechanical and Aerospace Engineering,*
*Cornell University, Ithaca, New York, USA*
jingjieyeo@cornell.edu

**[Abstract]** Amorphous silica (a-SiO$_2$) is a widely used inorganic material. Interestingly, the relationship between the local atomic structures of a-SiO$_2$ and their effects on ductility and fracture is seldom explored. Here, we combine large-scale molecular dynamics simulations and machine learning methods to examine the molecular deformations and fracture mechanisms of a-SiO$_2$. By quenching at high pressures, we demonstrate that densifying a-SiO$_2$ increases the ductility and toughness. Through theoretical analysis and simulation results, we find that changes in local bonding topologies greatly facilitate energy dissipation during plastic deformation, particularly if the coordination numbers decrease. The appearance of fracture can then be accurately located based on the spatial distribution of the atoms. We further observe that the static unstrained structure encodes the propensity for local atomic coordination to change during applied strain, hence a distinct connection can be made between the initial atomic configurations before loading and the final far-from-equilibrium atomic configurations upon fracture. These results are essential for understanding how atomic arrangements strongly influence the mechanical properties and structural features in amorphous solids and will be useful in atomistic design of functional materials.

## 1. Introduction

Silicon dioxide (i.e., silica or SiO$_2$) is a family of polymorphic materials that widely exists in nature as non-living substances or in various living organisms[Fernández et al., 2015; Flörke et al., 2008], and is frequently used for engineering purposes. For instance, due to its unique physical and chemical properties, SiO$_2$ is currently the main raw material for glass['Application of Silica Glasses in Practice', 1991; Kotz et al., 2017] and concrete[Mehta and Ashish, 2020; P. P. et al., 2021], and is also commonly used in the food[EFSA Panel on Food Additives and Nutrient Sources added to Food (ANS) et al., 2018] and cosmetics[Fytianos et al., 2020] industries. There are two major structures of silica: crystalline and amorphous. In this study, we mainly focus on amorphous silica (a-

$SiO_2$). Unlike crystalline silica which is predominantly found in nature and constitutes more than 10% by mass of the Earth's crust, amorphous silica is mostly man-made and manufactured in huge quantities for daily use[Croissant et al., 2020]. Due to its unique properties such as good mechanical strength, high dielectric strength, formability, and chemical durability, amorphous silica is an important material with a broad range of applications, such as semiconductor chips, window panels, telescope glasses, and optical fibers. However, like other oxide glasses, a-$SiO_2$ is susceptible to brittle fracture, especially in solids with surface flaws[Wondraczek et al., 2011], which in turn prohibits some engineering applications.[Yoshida, 2019] Hence, gaining a deeper understanding of its deformation and fracture mechanisms is of crucial importance.

Various mechanisms of fracture for a wide variety of material properties, dynamic responses, and structures were studied with multiscale computational modeling methods[Elliott, 2011; Fish et al., 2021; Horstemeyer, 2009; Peng et al., 2021; van der Giessen et al., 2020]. Brittle materials may rupture catastrophically within the elastic regime, whereas ductile materials may have high fracture toughness through irreversible deformation[Pineau et al., 2016; Tang et al., 2021; Xi et al., 2005]. Notably, amorphous materials like a-$SiO_2$ in our study may exhibit brittle-to-ductile transitions that can be driven by densification[Yuan and Huang, 2015], nanoscale sample sizes[Luo et al., 2016], nanoparticles consolidation[Zhang et al., 2019], or electron-beam shaping[Zheng et al., 2010]. Due to significant differences compared to the deformation and fracture mechanisms of crystalline solids which has been extensively studied, it still remains very challenging to determine the origins of ductility and fracture propensity of a-$SiO_2$ and other non-crystalline materials[Du et al., 2021; Tang et al., 2021; Wang et al., 2016].

Previous studies proposed the concept of shear transformation zones (STZs)[Argon, 1979; Ding et al., 2014; Falk and Langer, 1998] which act as the carriers of ductility in amorphous materials and are controlled by multiple factors[Cao et al., 2018; Cubuk et al., 2017; Patinet et al., 2016; Xu et al., 2018]. However, the STZs are most commonly studied in metallic glasses rather than oxide glasses and the brittle nature of oxide glasses muddies the existence of such plastically deforming regions. Recently, the advent of machine learning (ML) methods has advanced new ways of studying the mechanics within materials[Butler et al., 2018; Hsu et al., 2020; Qin et al., 2020; Rahman et al., 2021; Xie and Grossman, 2018; K. Yang et al., 2019; Z. Yang and Buehler, 2022], and several data-driven models were proposed to capture the deformation mechanism of amorphous solids and predict the propensity of fracture[Cubuk et al., 2015; Du et al., 2021; Fan and Ma, 2021]. Through these applications of ML methods for studying fracture propensity, a structural metric termed "softness" was introduced to describe an atom's susceptibility to rearrangement. Here, we apply a more direct and intuitive way to depict the local structural environment and decode the relationship between fracture behaviors of a-$SiO_2$ and its initial static structure using ML methods.

In our study, we focus on two major tasks: (1) studying the deformation and fracture mechanisms of a-$SiO_2$ at the nanoscale, and (2) predicting the fracture propensity of a-$SiO_2$ by mapping the initial atomic configurations to their future structural state(s). To do this, we prepare a series of different pressure-quenched samples and perform numerous molecular dynamics (MD) simulations to deform the samples under uniaxial tension. We further analyze the simulation results through the Open Visualization Tool (OVITO) and Visual Molecular Dynamics (VMD) and explore the potential relationships between the increase in macroscopic ductility and the local bond-switching events. We also construct a classification-based machine learning model using three different types of algorithms to predict the bond-switching activities upon fracture given the initial structure, which will help us in understanding the fracture propensity based on atomistic origins. The schematic of our study herein is shown in Figure 1.

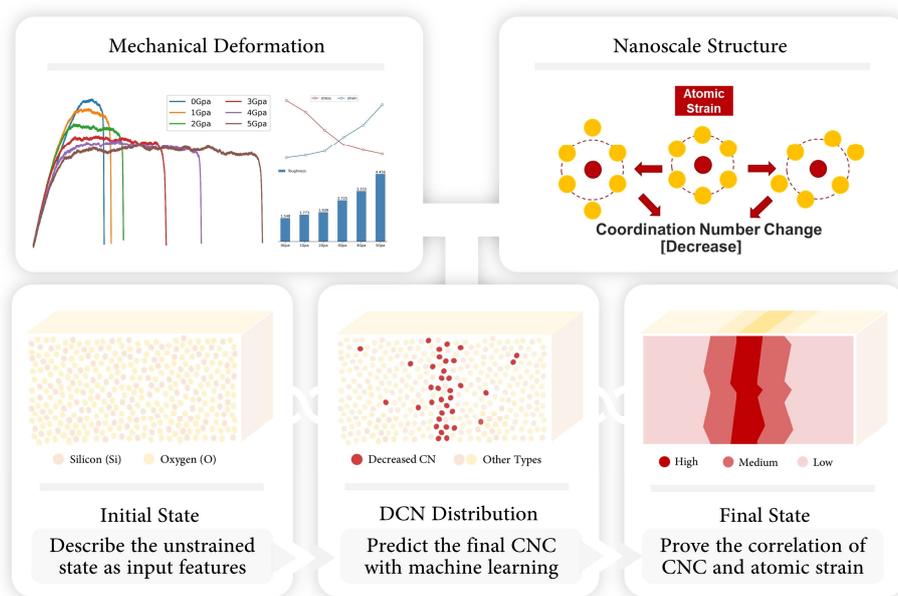

**Figure 1 Schematic of our numerical experiments.** We explored the correlation between the mechanical deformation and the atomistic structures based on molecular dynamics modelling and simulations of a-SiO2. We predicted the fracture surface in the final state based on the initial unstrained configuration. The connection between simulations and data-driven analyses provided holistic insight in the mechanical properties of amorphous silica.

## 2. Results and Discussion

### 2.1. *Mechanical Response in Macroscale*

To study the fracture propensity of amorphous silica, a series of a-SiO$_2$ samples was deformed under uniaxial tension with MD simulations using LAMMPS. These samples were prepared with a melt-quenching process by applying a range of constant hydrostatic pressures (0, 1, 2, 3, 4 and 5 GPa) for densification. Through this process, we obtained diverse mechanical responses due to changes in the atomic coordination.

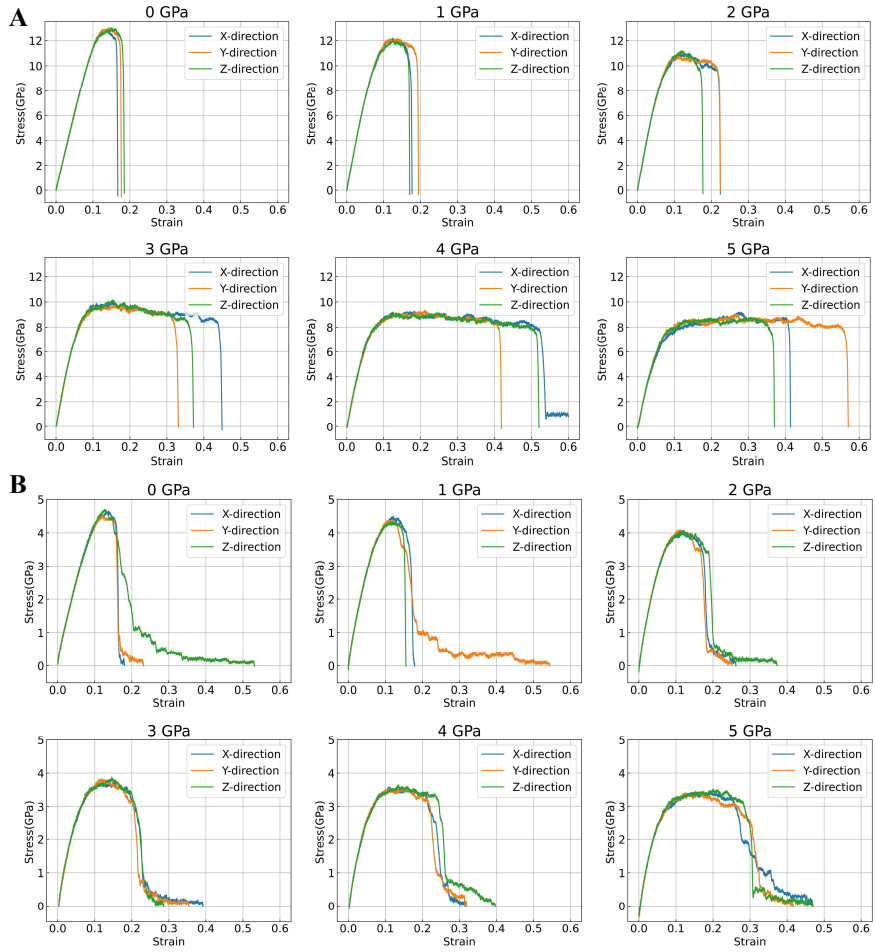

**Figure 2 Stress-strain curves under various conditions.** (A) Bulk amorphous silica with quenching pressures ranging from 0 to 5 GPa. (B) Amorphous silica nanowires with quenching pressures ranging from 0 to 5 GPa.

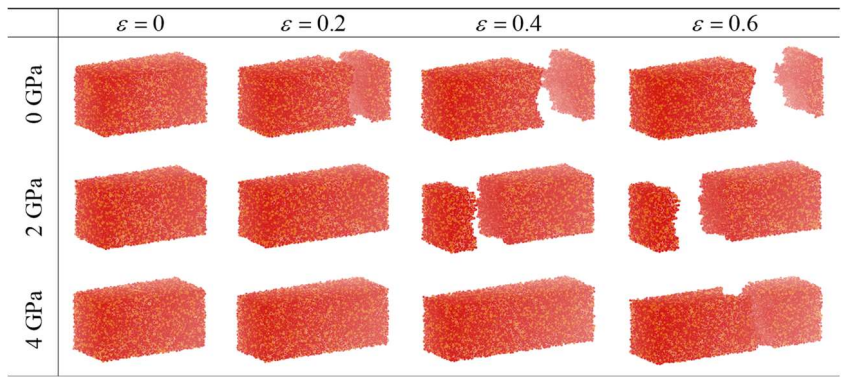

**Figure 3 Snapshots of the tensile simulation process.** The uniaxial tensile deformation of three different pressure-quenched (0 GPa, 2GPa, and 4 GPa) samples are shown at the applied strains of 0, 0.2, 0.4, and 0.6.

We obtained the stress-strain curves under tensile loading (Figure 2A) using quenching pressures ranging from 0 to 5 GPa, and each simulation was triplicated by loading in each Cartesian direction (x, y, and z) after the melt-quenching process (see Methods 3.1.1). Consistent with previous studies, the a-$SiO_2$ showed a clear brittle to ductile transition as the quenching pressure increased. When the quenching pressure was relatively low (< 2 GPa), the samples experienced a sudden drop after reaching the maximum stress (i.e., the material's yield strength) as the crack propagated instantaneously at a maximum strain of no more than 0.2. However, as the quenching pressure was increased, the maximum strain before fracture also increased, indicating a smooth transition into the regime of plastic deformations. For example, a sample quenched at 5 GPa might not fracture until > 40% uniaxial strain was attained, doubling that of lower-pressure quenched samples. We illustrated the atomic snapshots of such brittle to ductile transitions for a-$SiO_2$ densified at 0, 2, and 4 GPa (Figure 3) by extracting four critical moments until fracture at uniaxial strains of 0, 0.2, 0.4, and 0.6 to capture the deformation process. These illustrations agreed with our observations of fracture in the stress-strain curves, showing that the densification induced by pressure quenching can greatly increase the nano-ductility of a-$SiO_2$ samples, and the brittle-to-ductile phenomenon emerges as the pressure increased beyond 2 GPa. Typically, using the results in the y-direction as an example (see Figure 4A), as the quenching pressure rose from 0 GPa to 5 GPa, the maximum strain grew from around 0.18 to 0.58. This behavior is consistent with previous experiments and simulations[Du et al., 2021; Muralidharan et al., 2005; Pedone et al., 2008; Yuan and Huang, 2012, 2015], including other amorphous oxide glasses such as a-$Al_2O_3$. Contrary to the trend in maximum strain, the maximum stress (i.e., the ultimate tensile strength) of each sample decreased from 13 GPa to 9 GPa as the quenching pressure increased. This observation could be explained by the slight difference in the energy absorption and dissipation between brittle and ductile materials under deformation. a-$SiO_2$ samples quenched at higher pressures would be more densified and deformed mainly through shear flow[Yuan and Huang, 2015], hence yielding earlier and quickly transitioning into the plastic regime. With lower quenching pressures, yielding occurred at higher strains (~ 10% strain at 0 GPa) since the absorbed energy continued to increase the sample's elasticity instead of turning into plastic energy. Consequently, yielding at smaller strains resulted in lower ultimate tensile strengths, hence samples quenched at higher pressures usually had lower maximum stresses (Figure 4B).

Toughness is the ability of a material to absorb energy and deform without fracture. Enhancing toughness requires both increased strength and ductility. From our results, higher quenching pressure resulted in higher maximum strains and lower maximum stresses, but these two factors had confounding effects on the material's toughness. Toughness is equal to the amount of absorbed energy per unit volume prior to fracture and mathematically described as[Askeland and Wright, 2016]:

$$\text{Toughness} = \frac{\text{energy}}{\text{volume}} = \int_0^{\varepsilon_f} \sigma d\varepsilon \qquad (1.1)$$

where $\sigma$ and $\varepsilon$ represent stress and strain, and $\varepsilon_f$ is the strain upon rupture. Therefore, the toughness for different pressure-quenched samples could be obtained by integrating the areas under the stress-strain curves (Figure 4D). As quenching pressure increased from 0 to 5 GPa, the toughness for each sample was 1.548, 1.773, 1.939, 2.725, 3.333, and 4.456 GJ/m$^3$, respectively, increasing monotonically in a similar manner as the maximum strains. Therefore, even though more densified a-SiO$_2$ at higher quenching pressures had poorer mechanical strengths, the enhanced ductility dominated the energy absorption process and led to higher toughness for the material. Furthermore, the higher toughness is closely related to higher fracture energy which will be discussed further in Section 2.2.

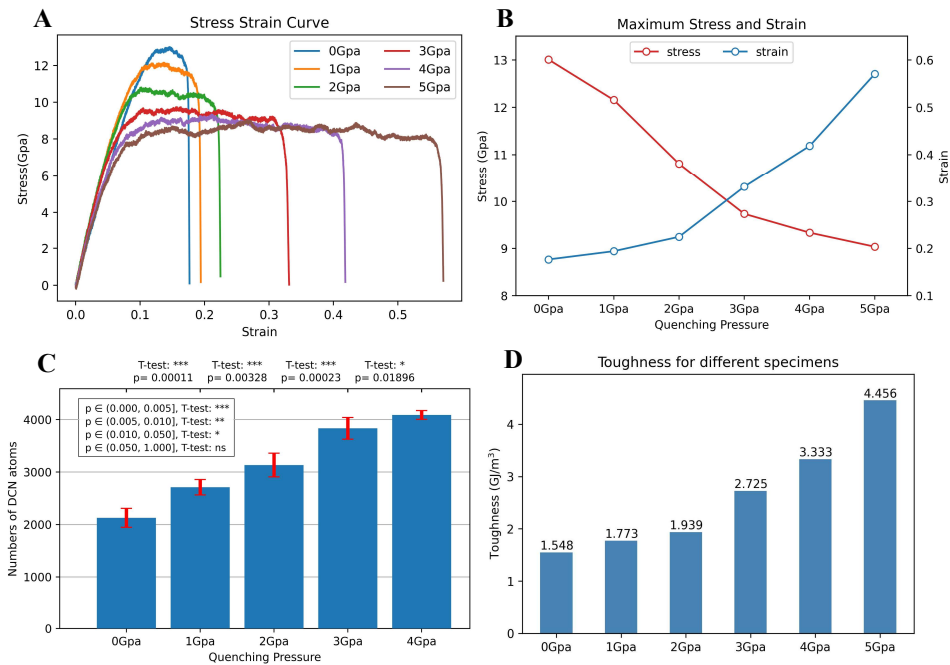

**Figure 4 Trends in stresses, strains, atomistic structure, and toughness as the quenching pressure increased.** (A) The six stress-strain curves are combined and plotted in a single figure. (B) The trend of the maximum stresses and strains of the six pressure-quenched samples. (C) The average numbers of DCN atoms at the final state, and the statistical hypothesis test to check if it is significantly different between neighbour samples. (D) Toughness for different pressure-quenched samples.

Other than simulations conducted using fully periodic samples (i.e., no free surfaces), the same uniaxial tensile deformation was applied to a-SiO$_2$ nanowires with free surfaces in two orthogonal directions and periodic in the axial direction (see Methods 3.1.2). From their stress-strain curves in Figure 2B, these nanowires had the same trend in mechanical properties as the periodic samples did: a higher pressure-quenched sample was associated with a higher failure strain (maximum strain) and a lower ultimate tensile strength (maximum stress). However, a-SiO$_2$ nanowires had lower strength and ductility than fully periodic a-SiO$_2$ models. Within the same range of quenching pressures, the ultimate tensile strength for nanowires was between 3.5 to 4.9 GPa, while that of periodic structures ranged between 9.0 to 13.0 GPa. Additionally, the failure strains of each a-SiO$_2$ nanowire in the x-, y-, and z-directions are almost the same even if the quenching pressure is relatively high, suggesting the dominance of free surface effects on yielding. Furthermore, the fracture process of an a-SiO$_2$ nanowire was milder and more progressive, attributable to necking during deformation of a-SiO$_2$ nanowires. Once necking began, stresses started to concentrate at the necking region and plastically deformed until fracture. The necking region, instead of the whole sample, withstood most of the localized stresses and plastic deformations, which led to comparably lower failure strains and ultimate tensile strengths.

To bridge the connections between atomic configurations in their initial states and fracture states, we next performed statistical analyses of the atomic configurations and trained an ML model to predict fracture from initial atomic configurations.

### 2.2. *Mechanical Response in Microscale*

Bond-switching frequently occurred throughout the deformation process, thereby reorganizing the local structures around the atoms and their atomic coordination. Hence, we would like to monitor the bond-switching events quantitively to investigate whether such events affected the propensity to fracture. To analyze these events, we measured the coordination numbers (CN), which is the number of atoms bonded to a central atom[Gold, 2019] (see Methods section 3.2.3). We initiated the measurements upon applying uniaxial tensile deformation and we mostly observed 4-fold coordinated silicon (CN=4) and 2-fold coordinated oxygen (CN=2) in the initial, non-deformed states. With increasing strain, the local atomic environment could change in three ways: a decrease in coordination number (DCN), an increase in coordination number (ICN), or the coordination number remained unchanged (UCN). Therefore, we ascribed all atomic coordination number changes (CNC) to these three categories: DCN, ICN, and UCN.

To observe each atom's coordination environment, the total numbers of DCN and ICN events was measured every time the strain had increased by 0.05 (Figure 5A and B).

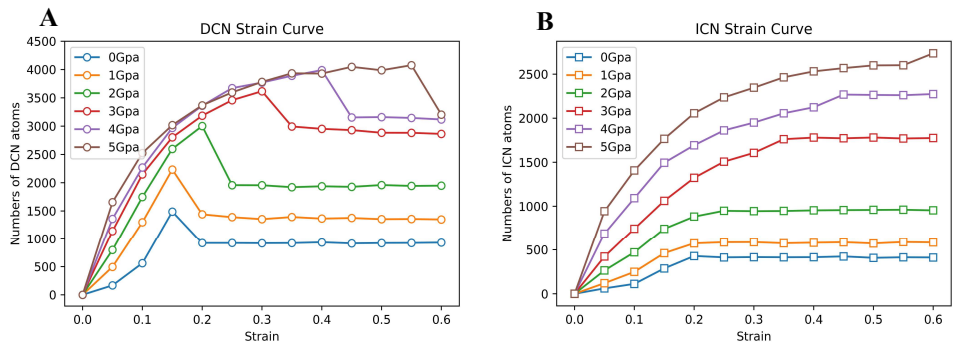

**Figure 5 The dynamics of coordination number change.** (A) Numbers of decreased CN atoms versus deforming strain for each sample. (B) Numbers of increased CN atoms versus deforming strain for each sample. All of them show a generally increase trend.

The numbers of atoms with either DCN or ICN generally increased in tandem with the applied strain, suggesting that more and more bond-switching events occurred as the a-SiO$_2$ samples were progressively deformed, thereby facilitating structural reorganization to dissipate the material's potential energy during fracture[To et al., 2021]. More densified samples had higher numbers of bond-switching events throughout the deformation process as well. With quenching pressure increasing from 0 to 5 GPa, the numbers of DCN and ICN atoms at the strain of 0.6 grew from 929 to 3204, and from 411 to 2736, respectively. These CNC events were also observed at fracture. By utilizing the data in multiple simulations and computing the statistical significance through p-values, we found that the average number of DCN atoms upon fracture grew significantly as the quenching pressure increased. These observations could help rationalize why samples quenched under higher pressure had better nano-ductility: bond-switching events delayed crack propagation by allowing the atomic structure of a-SiO$_2$ to reorganize before fracture finally occurred.

Although both DCN and ICN increased in tandem with increasing strain, there were still some differences between these two bond-switching activities. From Figure 5, all the ICN curves showed a consistent increase before attaining a steady state as strain increased from 0 to 0.6, while none of the DCN curves displayed this plateau. In fact, after continually increasing to the maximum amount, the number of DCN atoms decreased a little at the moment of fracture before plateauing. This observation can serve as a validation of Yuan et al.'s assumption[Yuan and Huang, 2015] which proposed that the deformation mechanism in amorphous solids is similar to that in polymers (untwining network chains and then stretching) as densified a-SiO$_2$ also had a two modes of deformation mechanism: shear flow and bond stretching. The decreasing portion of the DCN curves at the point of fracture is mainly caused by bond stretching during deformation. Once the sample fractured into two, some of the stretched bonds break while some of the others are not stretched anymore, resulting in a sudden drop in the number of DCN atoms. However, the DCN curves do not drop to zero due to the shear flow mechanism, which reorganizes the local structure plastically so that the decreased CN state will not recover even after fracture.

From Figure 5, under the same quenching pressure, atoms with decreased CN always outnumber atoms with increased CN at every strain that we considered. Combining the previous discussion on the deformation mechanism, we infer that DCN events may be more sensitive to how cracks originate and propagate between the two bond-switching activities, and therefore may be the key factor in analyzing fracture propensity. The reason why we choose DCN as a predictive metric can also be understood from the nanoscopic relationship between DCN and atomic strain, and the spatial distribution of this bond-switching activity (Figure 5). This is discussed in Section 2.3 and we relate DCN and the position where fracture occurs.

### 2.3. *Linking CNC with Crack Position*

We use the atomic strain as our metric for fracture as fracture only occurs at locations with maximum atomic strain compared with other locations in the solid. This can be understood from both theoretical means and our simulation results. We know that cracks exist in the regions where atoms are mostly separated (i.e., where the atoms' nonaffine displacements are high), and the atomic strain tensors are calculated exactly from averaging the discrete particle displacements over finite volumes (see Method 3.2.2). Hence, the position of the crack can be observed from the atomic strain. The same conclusion can also be obtained from our simulations. In Figure 6, using a sample quenched at 2 GPa as an example, the atoms are colored from dark purple to light yellow as atomic strain increased. With this color scheme, the fracture surface was eventually blanketed by atoms in yellow. This phenomenon could also be seen in samples stretched in other directions and quenched at other pressures. Hence, both from theoretical calculations and simulation results, the location of fracture corresponded to the areas with maximum atomic strain.

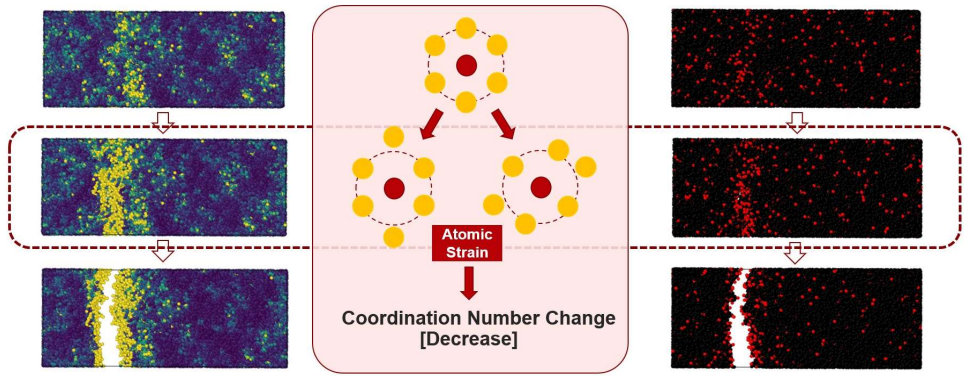

**Figure 6 Illustration of the relationship between atomic strain and DCN.** The left column shows how atomic strain varies during the deformation, while the right column shows the spatial distribution of DCN atoms in the same configuration. The central portion illustrates how atomic strain induced changes in the coordination number.

Using this observation, we can relate DCN and the atomic strain. In Figure 6, we illustrate the possible bond-switching activities of a 6-fold atom under normal strain and shear strain. The dashed line surrounding the central atom denotes the cutoff circumference which determines which neighboring atoms should be included when calculating the CN. Both normal and shear atomic strains induced changes in coordination environment: if the atomic strain was sufficiently high, the atom would possibly have a decrease in its coordination number, which resulted in the DCN activity in the sample that not only occurred in 6-fold atomic coordination, but also in other atoms with CNs equal to 5, 4, or 3. To confirm this observation with our simulation results, from Figure 6, we found that atoms with high atomic strains and atoms with decreased CN largely resided in the same regions. This was extremely pronounced in the vicinity of the eventual fracture surface, where both highly strained atoms and DCN atoms aggregated.

To quantitively analyze the connection between DCN and crack positions (indicated by atomic strain), we analyzed the final state upon fracture when the number of DCN atoms was maximum. Notably, for the model's simplicity and significance, we projected the atoms onto the direction where the uniaxial tensile test was conducted to convert the model from 3D to 1D. The distribution functions for DCN atoms and atomic strain (volumetric strain or shear strain) could then be calculated straightforwardly. For the DCN distribution, we directly considered the number of atoms with decreased CN per unit length. For strain distribution, we calculated the average value of atomic strain for all the atoms within a unit length. The unit length was set as 0.02 and all the values were calculated in steps of 0.001 between the normalized length of 0 to 1. Figure 7A illustrates these calculations using a-$SiO_2$ quenched at 0 GPa. The maxima of both the strain and DCN distribution corresponded to the precise position of crack initiation, while the distribution of highly strained atoms and atoms with the highest DCN were mostly located in the same region, quantitively demonstrating the strong correlation between DCN and atomic strain, where the predicted positions of fracture initiation were 0.778 and 0.772 respectively. Samples quenched at 1, 2, 3, and 4 GPa also displayed this strong correlation between strain and DCN (Figure 7B), demonstrating that high atomic strain was closely associated with DCN activity. Therefore, the eventual position of crack initiation could be predicted by the spatial distribution of the DCN atoms.

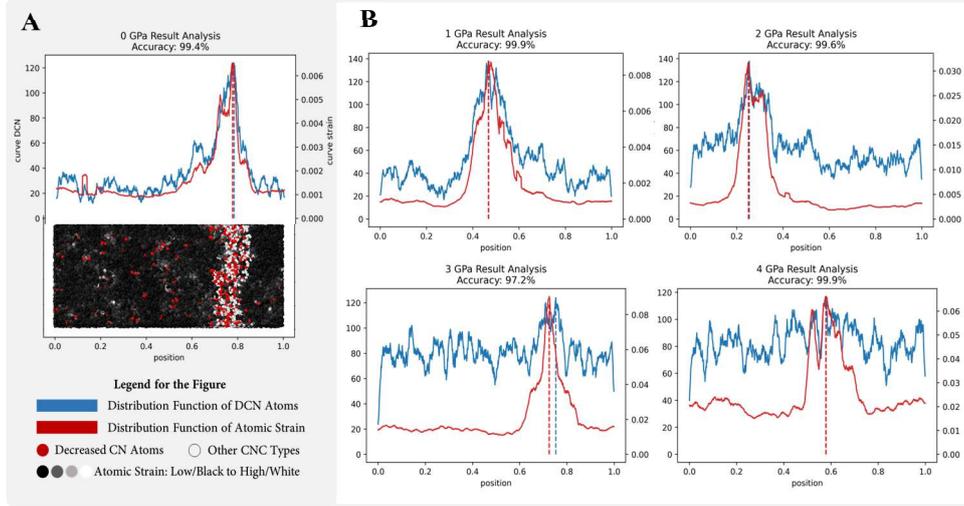

**Figure 7 Quantitively analysis for atomic strain and DCN.** (A) An illustration of how the distribution functions were generated, using a sample quenched at 0 GPa as an example. (B) The distribution functions of DCN atoms and atomic strain for 1 to 4 GPa samples, respectively. Both the DCN and highly-strained atoms mostly overlapped in the same region, and the predicted and the exact fracture positions were almost identical.

### 2.4. *Linking Initial Atomic Environment with CNC*

To demonstrate the utility of our findings, we predicted the changes in CN upon fracture simply by having information on the atomic configurations at the initial, unstressed state using machine-learning (ML) to train a classifier that accurately predicted the CNC. As defined above, the initial atomic configuration was taken right after the system was equilibrated before applying tensile deformation, while the final state upon fracture was when the failure strain was attained and the system had the maximum number of DCN atoms. We treated these two states as the beginning and the end of the uniaxial tensile deformation, and the output labels of the ML classifier (i.e., DCN, ICN, or UCN) could be determined by comparing the CN between these two states. For numerical simplicity, we assigned the values of -1, 1, and 0 to the labels of DCN, ICN, and UCN respectively. The input features were families of structure functions containing 47 radial structure functions (RSFs) and 22 angular structure functions (ASFs) that were computed based on the initial atomic configurations (see Methods 3.2.4). The data was selected and prepared for the ML classifier using procedures outlined in Methods 3.3.1, and the binary ML classifiers (DCN-ICN and DCN-UCN) were then trained and tested.

### 2.4.1. *RBF SVM Classification with RSF as the input feature*

For baseline comparison, the supervised learning algorithm of support vector machine (SVM) was used to train our ML classifier. Each training data that needed to be classified was expressed as a point in an N-dimensional space, and SVM performed the classification task by finding a hyperplane that mostly separated the two categories (i.e., maximized the width of the gap between the two categories). Only the 47 radial structure functions were used as inputs and the RBF kernel was applied to further map these inputs into a higher dimensional feature space. After confirming the hyperplane using the training dataset, new examples from the testing dataset were mapped into the same space and classified based on which side of the hyperplane they fell on.

The classification model was trained for all the samples quenched at 0 to 4 GPa. We omitted the data from the sample quenched at 5 GPa which did not show a significant difference in terms of the maximum number of DCN atoms compared with the 4 GPa quenched sample when applying the statistical hypothesis test (i.e., t-test), and the DCN atoms at the final state were distributed quite evenly throughout the sample, making it hard to distinguish the location of fracture. Figures 8A and B depict the accuracies for both the training and testing sets, as well as the best hyper-parameters, for the sample quenched at 2 GPa. The bar plots in Figures 8C and D show the final testing accuracies for the DCN-ICN classifiers and DCN-UCN classifiers for all the samples, and both classifiers performed quite well. For the DCN-ICN classifier, all the samples achieved an accuracy greater than 89.52% with the highest reaching 92.57% (0 GPa), while for the DCN-UCN classifier, the results were almost the same with the highest accuracy of 91.43% (1 GPa). With this RBF SVM result as a baseline, we compared the accuracies of other ML algorithms and the effects of including both RSF and ASF in the input features.

### 2.4.2. *Broader sets of ML Classifiers with both RSF and ASF as input features*

The seven ML algorithms used here can be divided into three major types: Nearest Neighbors, Support Vector Machines, and Trees and Forest. We first focused on the DCN-ICN classification in the 2 GPa quenched sample. With Nearest Neighbors, we adopted the *k*-nearest neighbors (KNN) algorithm where an object from the testing set is classified by a plurality vote of its neighbors which come from the training set. Specifically, the number of neighbors that we considered is given by the parameter *k*, and by calculating the Euclidean distances, the *k* nearest neighbors is selected and the new testing object is assigned to the majority category among them. The default value for *k* is 3 and we obtained accuracies of 52.86% (before data normalization) and 84.29% (after data normalization, see Method 3.3.1). Normalization significantly improved the accuracy as KNN is a distance-based algorithm and normalizing the data can avoid vastly different scales in input features[Piryonesi and El-Diraby, 2020]. We further tuned the hyper-parameter *k* via grid search[Goodfellow et al., 2016] and chose the best *k* value of 25 by comparing the cross-validation results (Figure 9A), which means that the 25 nearest neighbors were considered every time we assigned a new object, and the prediction accuracy was further improved to a high of 90.95%.

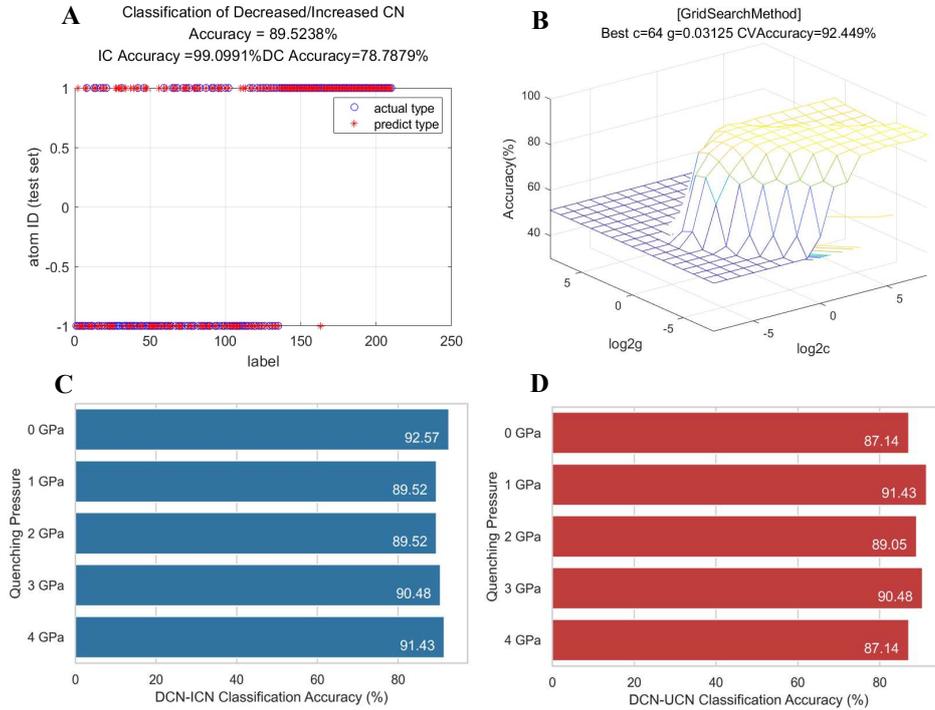

**Figure 8 Data, accuracy, hyper-parameters, and results using the RBF SVM.** (A) The testing accuracy of DCN-ICN classifier using the 2 GPa quenched sample. (B) The grid search map when deciding the best hyper-parameters for RBF SVM. (C) DCN-ICN classification accuracies of all the samples under different pressures. (D) DCN-UCN classification accuracies of all the samples under different pressures.

With Support Vector Machines, besides applying the RBF SVM, we also applied two other algorithms of linear SVM and polynomial SVM. In linear SVM, no kernel is used, and finding the hyperplane can be transferred into optimizing the trade-off parameter C to balance larger margin size and higher correctness in classification. After tuning the hyperparameters, we obtain C=0.25and the testing accuracy was 86.19%. With the polynomial SVM, the hyperplane is constructed in a transformed feature space with a corresponding kernel function, which shares a similar logic as RBF SVM. The accuracies for classifiers trained by polynomial SVM and RBF SVM were 85.24% and 90.95%. Notably, with the same training and testing objects, the accuracy using RBF SVM slightly increased from 89.52% to 90.95% after we included the ASF in the feature space. This signified that angular structure functions could help to describe the local atomic environment, but with only minor effects as most information might already be contained in the RSF.

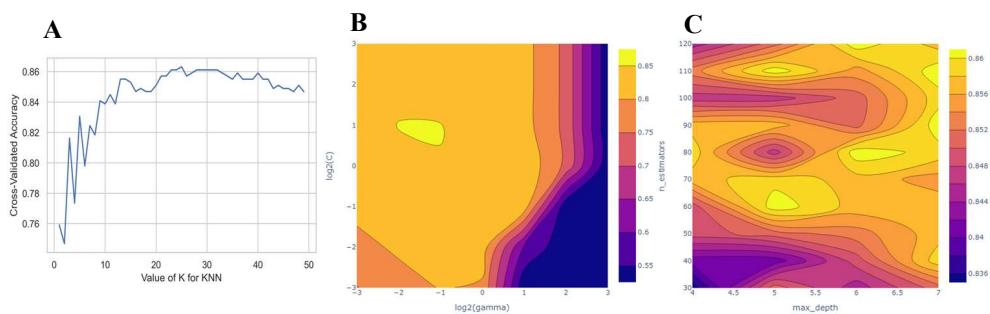

**Figure 9 Hyper-parameter tuning process.** The hyper-parameters were tuned via grid search method for (A) KNN; (B) RBF SVM; (C) Random Forest where all the parameters were tested exhaustively and the best set was determined based on the cross-validation results.

With Trees and Forest, another three algorithms, namely decision tree (DT), random forest (RF), and extra trees (ET), were applied to train the classifier. A decision tree is constructed by learning simple decision rules inferred from the data features. To look for the best split from the top down, we quantified and compared the impurity using a node-splitting function called "Gini Impurity" instead of "Entropy and Information Gain", according to our cross-validation results. We also found that it was optimal to set the maximum depth of the tree as 4 and consider all the 69 input features. Based on these settings, the prediction accuracy of the DCN-ICN classifier was 82.38%. This accuracy was not as high as before since DT was not flexible when it comes to classifying new samples[Hastie et al., 2009]. RF can improve the classification as RF consists of multiple decision trees and generates the prediction results voted by all these trees. By using a bootstrapped dataset[Breiman, 2001] and only considering a random subset of features at each step[Tin Kam Ho, 1995], we attained a higher accuracy of 90.48% after tuning the hyper-parameters as max_depth=6 and n_estimators=120. Here, max_depth was the maximum depth of the tree, and n_estimators was the number of trees in the forest. Extra trees (i.e., extremely randomized trees) is similar to RF but chooses the splitting feature randomly instead of the optimum one[Geurts et al., 2006]. ET usually reduces variance at the cost of a slight increase in bias. From our results, the prediction accuracy using ET was 90%.

By adopting three major types of ML classification algorithms together with RSF and ASF as input features, 4 out of the 7 algorithms achieved an accuracy of no less than 90% (Figure 10A). The accuracies using the original data and normalized data are shown in Figure 10B together with the accuracies after hyper-parameter tuning via grid search. All the distance-based algorithms (KNN, SVM) had a clear increase in accuracy after data normalization while all the tree-based algorithms (DT, RF, ET) remained similar. This could be attributed to the fact that the former uses distances between data points to evaluate their similarity, but the latter only splits a node based on a single feature which is not affected by the range of inputs. Moreover, most of the algorithms performed better after conducting grid search to select the best hyper-parameters.

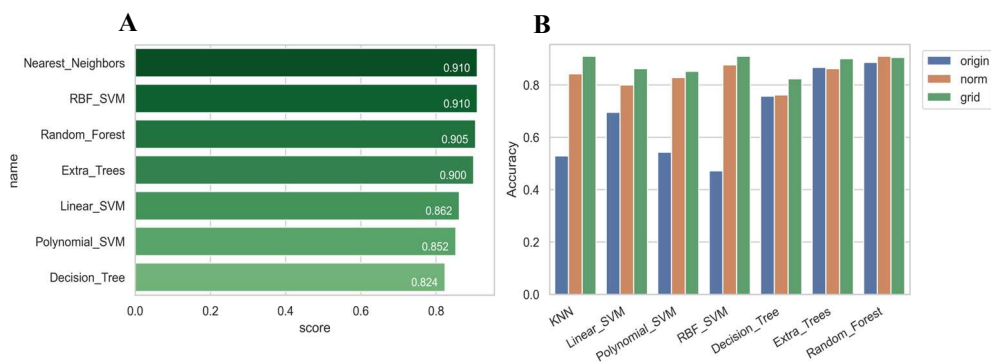

**Figure 10 Results of multiple ML classifiers.** (A) Accuracies of the 2 GPa DCN-ICN classifier trained by seven different machine learning algorithms in three major types. (B) The differences between using the original data and the normalized data, and the differences between with and without hyper-parameter tunning.

The same level of accuracy was not only seen in the DCN-ICN classifier using 2GPa quenched sample, but also in the DCN-UCN classifier and other pressure-quenched samples. Therefore, based on the performance of these machine learning methods, we successfully built the connection between the local atomic environment at the initial state and the coordination number change (CNC) at the final state.

## 3. Methods

### 3.1. *Molecular Dynamics Simulation*

#### 3.1.1. *Amorphous Phase Construction*

We used the Large-scale Atomic / Molecular Massively Parallel Simulator (LAMMPS) software to perform all MD simulations. The Vashishta potential was used to model the interatomic interactions, which can be written as:

$$U_i = \frac{1}{2}V_2(r_{ij}) + \frac{1}{2}\sum_{j \neq i}\sum_{k \neq i,j} h_{ijk} \qquad (3.1)$$

where $V_2(r_{ij})$ is the two-body part pairwise potential and $h_{ijk}$ is the modified form of the three-body part of the Stillinger-Weber potential. We applied a Nose-Hoover thermostat and barostat to control the temperature and pressure and used fully periodic boundary conditions (PBC) in all three Cartesian directions. To generate a block of amorphous silica (a-$SiO_2$), we first constructed an initial unit of a building block. The unit contained 9 atoms (3 Si and 6 O) and this unit block was replicated 15 times in all three Cartesian directions to generated a larger crystalline block that had 30,375 atoms.

To obtain amorphous silica, we applied a quenching procedure, which will be specified in the following content. Many articles had specified how to generate amorphous materials via quenching[Du et al., 2021; Frankberg, Kalikka, García Ferré, et al., 2019; Gutiérrez and Johansson, 2002], and here is our process: we first started with a phase in the NVT ensemble at 5000 K for 45 ps to melt silica and randomize the initial crystalline structure. Then, we cooled the system down to 3000 K under a pressure of 100 atm, after which the pressure was raised to different certain values (0,1,2,3,4 and 5 GPa) in preparation for quenching while the temperature was maintained at 3000 K during these 60 ps. After equilibrating for an additional 10 ps, the system was cooled to 300 K in the NPT ensemble under the pressure that has been set, which lasted for 650 ps. During this period, with the temperature going lower, the mobility of the atoms decreased and the silica system finally solidified. We then gradually decreased the quenching pressure to zero and relaxed the system at 300 K in the NPT ensemble for 10 ps. The timestep for the whole process was 0.001 ps and we output the thermal properties every 1000 steps. Through this melt-quenching process, we controlled pressure as our only variable, and samples were quenched independently from random liquid states to amorphous solids.

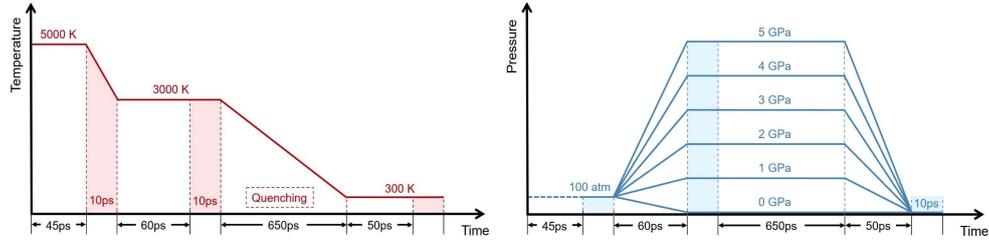

**Figure 11 Schematic of the entire melting and quenching process**

### 3.1.2. *Fracture Simulations*

To study the fracture propensity of a-SiO$_2$, we applied uniaxial tensile deformations. Before doing so, we duplicated the as-quenched cubic a-SiO$_2$ two times in one direction to ensure that we could apply the tensile simulation to the longer direction. Specifically, for each as-quenched sample, we duplicated three different times in the x, y, and z directions respectively, generating three new samples with 60,750 atoms in each of these systems, following similar methods published previously[Yuan and Huang, 2012] to generate larger systems. By doing so, we reduced the computational time and resources for quenching every sample from the liquid state and could conveniently observe differences and similarities with more samples when needed.

To ensure the robustness of our models, we considered two additional steps which would generate slightly different samples for analysis. a) Tempered a-SiO$_2$: we heated the sample to 3000 K and cooled it under the same pressure that we applied to it when quenching to create a tempered a-SiO$_2$. b) Nanowire a-SiO$_2$: we expanded the simulation box size in the directions that were orthogonal to the applied tensile deformation to create

a-SiO$_2$ nanowires with free surfaces. For tempered a-SiO$_2$, no significant differences were found, while for nanowire a-SiO$_2$, there were some dissimilarities (see Results 2.1).

Each system was finally relaxed for 10 ps at 300 K. As the fracture behaviors generated by different loading conditions normally will not be the same[Frankberg, Kalikka, Ferré, et al., 2019; Yuan and Huang, 2012], here we focused on the "plane stress" condition in our uniaxial tensile deformation: we chose the longer direction as our deforming direction and set the strain rate of $10^9$ s$^{-1}$. The simulation box was deformed for 600 ps, leading to a maximum strain of 0.6. At the same time, the lateral boundaries were controlled using the NPT ensemble to zero pressure while the temperature was maintained at 300 K. The strain and stress values were defined before the tensile deformation and output every 1000 steps for further inspection and analysis.

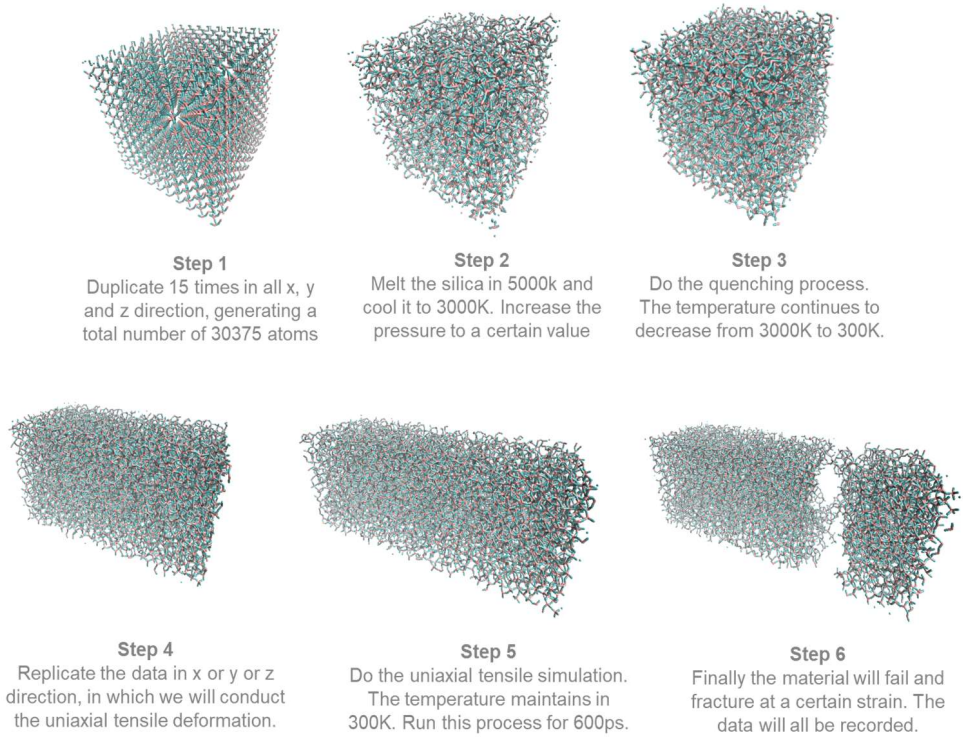

**Figure 12 Uniaxial Tensile Simulation**

### 3.2. *Features Generation*

3.2.1. *Displacement Vector*

The displacement vector is one of the key variables that can be calculated when from the simulation trajectory. Since we wanted to study the relationship between the initial undeformed state and the moment of fracture when the crack propagated, we defined the reference and current configuration respectively using the Open Visualization Tool (OVITO). We focused on the internal displacements of the particles, so we set the affine mapping to reference. By doing so, the particle positions at any points in time were remapped onto the reference configuration cell before calculating the displacement vectors, so that the macroscopic deformation was filtered out and we could focus on the non-affine displacements of the particles by subtracting the vectors between these two configurations.

Prior work[Falk and Langer, 1998] also proposed the concepts of cumulative nonaffine displacement $D$ and nonaffine square displacement $D^2_{min}$ with the same intention of quantitatively measuring the degree of local reorganization as follows:

$$D = \sum_{s=1}^{n} \sqrt{\Delta D^2_{s,min}} \tag{3.2}$$

where $\Delta D^2_{s,min}$ is the incremental nonaffine square displacement. However, since our model will only include the basic displacement vector rather than $D$ and $D^2_{min}$, we will not implement these calculations here.

### 3.2.2. *Atomic Strain*

We calculated the atomic strain tensor at each Si and O atom from the relative motion of its neighbors. Specifically, the calculation was conducted within the following steps. First, the atomic deformation gradient tensor $\boldsymbol{F}$ was calculated using the displacement vector that was mentioned before. Then, the Green-Lagrangian strain tensor $\boldsymbol{E}$ was derived for each atom[Kaye et al., 1998; Lubliner, 2008]:

$$\boldsymbol{F} = \boldsymbol{I} + \nabla \boldsymbol{u} \tag{3.3}$$

$$\boldsymbol{E} = \frac{1}{2}(\boldsymbol{F}^T \boldsymbol{F} - \boldsymbol{I}) \tag{3.4}$$

To output the atomic strains, we further calculated the volumetric strain from the tensor and the von Mises local shear invariant[Shimizu et al., 2007]:

$$E_{volume} = (E_{xx} + E_{yy} + E_{zz})/3 \tag{3.5}$$

$$E_{shear} = \left[ E_{xy}^2 + E_{xz}^2 + E_{yz}^2 + 1/6\left( \left(E_{xx} - E_{yy}\right)^2 + \left(E_{xx} - E_{zz}\right)^2 + \left(E_{yy} - E_{zz}\right)^2 \right) \right]^{1/2} \tag{3.6}$$

These two values could be calculated for each particle and they measure the unit change in volume and the shear deformations. They also play an important role in crack recognition as in our results.

### 3.2.3. Coordination Number

Coordination Number (CN) can be determined by counting the neighboring atoms to which the central atom is bonded. For crystalline silica, the common coordination numbers are 4 (for Si) and 2 (for O). For amorphous silica, however, the situation may be slightly different.

Radial pair distribution functions (RDFs) were first calculated before we determined the coordination numbers for a-SiO$_2$. It was a function of distance $r$ and measured the possibility of finding another atom from the central atom at this distance. For our a-SiO$_2$ samples, there were three kinds of atomic pairs, namely Si-O, Si-Si, O-O, but we focus only on Si-O. We set the cutoffs to be the distances at the bottom after the first peaks in the RDFs, which meant that atoms were considered bonded within this distance. Based on the data shown in figure 13A, the cutoff distance selected for our research was 1.8 Å.

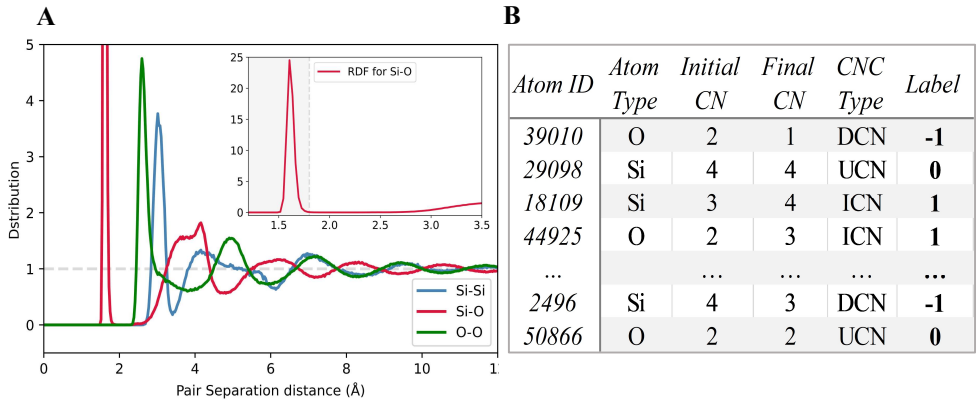

**Figure 13 Radial distribution functions and labelling method**

After setting the cutoff distance, the CN could be conveniently calculated by counting the number of atoms within the distance. However, as a single value of CN can only reflect a state at one moment, we further introduced the time domain and investigated the bond-switching activities during the whole tensile deformation until fracture occurred. The CNs from the initial unstrained state and the final state when the sample was about to crack were extracted for comparison. To quantify the bond-switching activities during this process, we considered a similar labeling method that was used for amorphous alumina[Du et al., 2021]: the atoms were labeled as DCN if the CN has decreased compared with the initial state or labeled as ICN if the CN has increased. If the CN remained unchanged, then we labeled them as UCN. With this method, every atom was labeled in terms of the changes in their coordination numbers and the neighborhood environment could be captured effectively during the tensile deformation.

### 3.2.4. *Structure Functions*

There are many ways to describe the atomic structure of a solid. Here, we use a method developed previously[Behler and Parrinello, 2007; Cubuk et al., 2015; Du et al., 2021] to depict the local structure of a-SiO$_2$ as well as generate the input features for the ML model. Specifically, for each particle $i$, we defined two families of structure functions: the radial structure functions (RSF) and the angular structure functions (ASF):

$$G_i(\mu) = \sum_j e^{-(R_{ij}-\mu)^2/L^2} \quad (3.7)$$

$$\psi_i(\xi,\lambda,\zeta) = \sum_j \sum_k e^{-(R_{ij}^2+R_{ik}^2+R_{jk}^2)/\xi^2}(1+\lambda\cos\theta_{ijk})^\zeta \quad (3.8)$$

Here in both structure functions, $R_{ij}$ represents the distance between particle $i$ and $j$. In the RSF, $\mu$ is a value ranging from 0.6 to 10 Å with an increment of 0.2 Å, while $L$ is the window parameter which is selected to be 0.3 Å. In the ASF, $\theta_{ijk}$ stands for the angle between particles $i$, $j$, and $k$, and $\xi, \lambda, \zeta$ are a series of parameters, to which we can assign different combinations of values to characterize the angular environment of an atom. The values we used are listed below:

**Table 1 Parameters for Angular Structure Function**

| | 1 | 2 | 3 | 4 | 5 | 6 | 7 | 8 | 9 | 10 | 11 |
|---|---|---|---|---|---|---|---|---|---|---|---|
| $\xi$ | 14.63 | 14.63 | 14.64 | 14.64 | 2.554 | 2.554 | 2.554 | 2.554 | 1.648 | 1.648 | 1.204 |
| $\lambda$ | -1 | 1 | -1 | 1 | -1 | 1 | -1 | 1 | 1 | 1 | 1 |
| $\zeta$ | 1 | 1 | 2 | 2 | 1 | 1 | 2 | 2 | 1 | 2 | 1 |
| | 12 | 13 | 14 | 15 | 16 | 17 | 18 | 19 | 20 | 21 | 22 |
| $\xi$ | 1.204 | 1.204 | 1.204 | 0.903 | 0.903 | 0.903 | 0.903 | 0.695 | 0.695 | 0.695 | 0.695 |
| $\lambda$ | 1 | 1 | 1 | 1 | 1 | 1 | 1 | 1 | 1 | 1 | 1 |
| $\zeta$ | 2 | 4 | 16 | 1 | 2 | 4 | 16 | 1 | 2 | 4 | 16 |

To better describe the atomic structure, we need the RSF to characterize radial density properties as well as the ASF to characterize bond orientation properties. So, for each particle $I$, by varying the parameter $\mu$ from 0.6 to 10 Å, i.e., $\mu_n = (0.6+0.2n)$ Å, with $n = 0, 1, 2, ..., 47$, we could obtain 47 RSF. Also, by varying 22 groups of values for parameters $\xi, \lambda, \zeta$, we expect to get the same number of ASF. We combined the 47 radial structure functions and 22 angular structure functions to create an input vector $X_i$ with a total number of 69 input features:

$$\boldsymbol{X}_i = [\underbrace{G_i(\mu_1), G_i(\mu_2), ..., G_i(\mu_{47})}_{47 \text{ RSFs}}; \underbrace{\psi_i(\xi_1,\lambda_1,\zeta_1), ..., \psi_i(\xi_{22},\lambda_{22},\zeta_{22})}_{22 \text{ ASFs}}] \quad (3.9)$$

### 3.3. *Machine Learning Methodology*

#### 3.3.1. *Data Preparation*

To predict the changes in CN in the final state from the initial state via ML, we first need to prepare the features (inputs) and the labels (outputs) of our model. As described above, the structure functions are our input features. We extract the coordinates of each atom at the initial state and calculated the 47 RSFs and 22 ASFs. Since we had 60750 atoms in a sample, we obtained an input matrix of 60750×69. For the output labels, the change in CN from the initial state to the final state was used to generate a label vector of 60750×1. Labels of DCN, ICN, and UCN were assigned the values of -1, 1 and 0.

We also normalized the data using min-max feature scaling as ML algorithms are sensitive to the range of values in the features. For each column:

$$X' = \frac{X - X_{min}}{X_{max} - X_{min}} \tag{3.10}$$

As we need to distinguish DCN from ICN and UCN, we created two binary ML classifiers of DCN-ICN and DCN-UCN. To prevent bias in the results due to the imbalanced dataset, each of the classifiers contained an equal number of 350 DCN atoms and 350 ICN (or UCN) atoms.

To train the classifier and evaluate the performance, we randomly split the newly formed dataset into training data (70%) and testing data (30%). For the same sample, despite the different machine learning algorithms we used, the train/test split was identical so that the results could be compared.

#### 3.3.2. *Pre-test Verification*

Before training the ML model, we need to first verify if the structure descriptors (RSF and ASF) could indeed capture the neighborhood environment of each atom by analyzing whether the structure functions can successfully distinguish between Si and O atoms. Using the case of 2Gpa quenching pressure as an example, the input data was maintained the same as before, which was a 60750×69 matrix containing 47 RSFs and 22 ASFs. As for the output data, we used the labels of atom type: the atom was labeled as 1 if it was silicon (Si) while labeled as 0 if it was oxygen (O). After splitting the data into training and testing sets, we deployed linear SVM using the LIBSVM package[Chang and Lin, 2011] to train the classifier. The result of our pre-test is shown in the results section, from which verify that our descriptors worked well for our purposes.

#### 3.3.3. *Applying ML to model training*

On the basis of the structure descriptors (RSFs and ASFs) and the corresponding labels (DCN, ICN, or UCN), we then applied three main classes of ML algorithms using the scikit-learn package[Pedregosa et al., n.d.] to train the classifiers:

a) Nearest Neighbours (KNN);
b) Support-Vector Machine (Linear SVM, Polynomial SVM, RBF SVM);
c) Trees and Forest (Decision Tree, Extra Trees, Random Forest);

### 3.3.4. *Parameter Tunning via Grid Search*

In many ML algorithms, there are parameters that cannot be directly learned and need to be defined manually and passed to construct estimators. Here, we tuned these hyper-parameters using grid search to increase the accuracy of our predictive model. For each algorithm, we first manually specified a subset of hyper-parameter space. Then, we searched exhaustively through this space to find the best set of hyper-parameters. During this process, cross-validation was conducted and we used the average score of the validation sets to evaluate the hyper-parameter(s) every time.

For each machine learning algorithm that was mentioned above, the hyper-parameters that need to be confirmed are as follows:

Table 2 Hyper-parameters Tunning Space

| Algorithms | Hyper-parameter Space |
|---|---|
| KNN | n_neighbors = range(1,50) |
| Linear SVM | C = 2**range(-3,4) |
| Polynomial SVM | C = 2**range(-3,4), degree = range(2,5) |
| RBF SVM | C = 2**range(-3,4), gamma = 2**range(-3,4) |
| Decision Tree | criterion, max_features, max_depth = [3,4,5,6,7] |
| Extra Trees | n_estimators = range(30,130,10), max_depth = [4,5,6,7] |
| Random Forest | n_estimators = range(10,110,10), min_samples_split = [2,3,4,5] |

In Decision Tree, criterion was selected between "gini" and "entropy" while max_features was selected between "None" and "sqrt".

## 4. Conclusion

We examined the deformation and fracture process of amorphous silica and linked their present unstressed atomistic configurations to their ultimate fracture states. To do so, we performed uniaxial tensile deformation of different pressure-quenched a-SiO$_2$ samples using MD simulations and found that more densified samples were always associated with higher failure strains and toughnesses. This observation was consistent with that of many amorphous solids in general. We further discovered that this increase in macroscopic ductility could be attributed to the local bond-switching events in microscale, which allowed the atomistic structure to reorganize and facilitated the dissipation of potential energy in the material. By analyzing the changes in the number of bond-switching events during deformation, we found that the development of decreased CN activities was closely related to the deformation mechanisms and inferred that DCN was the key factor in learning the fracture propensity. We introduced atomic strain as the fracture indicator and demonstrated that the spatial distributions of DCN atoms and highly strained atoms were strongly correlated. Thus, the fracture position could be identified by DCN. We adopted classification-based machine learning algorithms and revealed that the long-term bond-switching dynamics of atoms could be predicted given the initial unstrained structure. Our results suggested that both DCN-ICN classifier and DCN-UCN classifier could achieve a prediction accuracy of around 90%, which implied that the DCN at the final state could be successfully distinguished based on the initial atomic environment. Therefore, the connection of initial structure-DCN-fracture position was established and the fracture propensity could be predicted based on the initial atomistic configuration of an unstrained a-$SiO_2$ sample. Based on these observations in our study, we further envision that more types of mechanical deformations can be studied with the help of machine learning, which will facilitate the decoding of the atomistic relationship between nanoscale structure and macroscale properties and help us in designing materials with specific mechanical properties.

**Acknowledgments**

J.Y. acknowledges support from the US National Science Foundation under awards CMMI-2038057, EFMA-2223785, and ITE-2236190, as well as the computational resources provided by the NSF Advanced Cyberinfrastructure Coordination Ecosystem: Services & Support program under award TG-BIO210063.

18. Fernández, L. D., Lara, E., and Mitchell, E. A. D. [2015]. Checklist, diversity and distribution of testate amoebae in Chile. *European Journal of Protistology*, *51*(5), 409–424. https://doi.org/10.1016/j.ejop.2015.07.001
19. Fish, J., Wagner, G. J., and Keten, S. [2021]. Mesoscopic and multiscale modelling in materials. *Nature Materials*, *20*(6), 774–786. https://doi.org/10.1038/s41563-020-00913-0
20. Flörke, O. W., Graetsch, H. A., Brunk, F., Benda, L., Paschen, S., Bergna, H. E., … Schiffmann, D. [2008]. Silica. In Wiley-VCH Verlag GmbH & Co. KGaA (Ed.), *Ullmann's Encyclopedia of Industrial Chemistry* (p. a23_583.pub3). https://doi.org/10.1002/14356007.a23_583.pub3
21. Frankberg, E. J., Kalikka, J., Ferré, F. G., Joly-Pottuz, L., Salminen, T., Hintikka, J., … Masenelli-Varlot, K. [2019]. *Highly ductile amorphous oxide at room temperature and high strain rate*. 7. https://doi.org/10.1126/science.aav1254
22. Frankberg, E. J., Kalikka, J., García Ferré, F., Joly-Pottuz, L., Salminen, T., Hintikka, J., … Masenelli-Varlot, K. [2019]. Highly ductile amorphous oxide at room temperature and high strain rate. *Science*, *366*(6467), 864–869. https://doi.org/10.1126/science.aav1254
23. Fytianos, G., Rahdar, A., and Kyzas, G. Z. [2020]. Nanomaterials in Cosmetics: Recent Updates. *Nanomaterials*, *10*(5), 979. https://doi.org/10.3390/nano10050979
24. Geurts, P., Ernst, D., and Wehenkel, L. [2006]. Extremely randomized trees. *Machine Learning*, *63*(1), 3–42. https://doi.org/10.1007/s10994-006-6226-1
25. Gold, V. (Ed.). [2019]. *The IUPAC Compendium of Chemical Terminology: The Gold Book* (4th ed.). https://doi.org/10.1351/goldbook
26. Goodfellow, I., Bengio, Y., and Courville, A. [2016]. *Deep learning*. Cambridge, Massachusetts: The MIT Press.
27. Gutiérrez, G., and Johansson, B. [2002]. Molecular dynamics study of structural properties of amorphous $Al_2O_3$. *Physical Review B*, *65*(10), 104202. https://doi.org/10.1103/PhysRevB.65.104202
28. Hastie, T., Tibshirani, R., and Friedman, J. [2009]. *The Elements of Statistical Learning*. https://doi.org/10.1007/978-0-387-84858-7
29. Horstemeyer, M. F. [2009]. Multiscale Modeling: A Review. In J. Leszczynski and M. K. Shukla (Eds.), *Practical Aspects of Computational Chemistry* (pp. 87–135). https://doi.org/10.1007/978-90-481-2687-3_4
30. Hsu, Y.-C., Yu, C.-H., and Buehler, M. J. [2020]. Using Deep Learning to Predict Fracture Patterns in Crystalline Solids. *Matter*, *3*(1), 197–211. https://doi.org/10.1016/j.matt.2020.04.019
31. Kaye, A., Stepto, R. F. T., Work, W. J., Alemán, J. V., and Malkin, A. Ya. [1998]. Definition of terms relating to the non-ultimate mechanical properties of polymers (Recommendations 1998). *Pure and Applied Chemistry*, *70*(3), 701–754. https://doi.org/10.1351/pac199870030701
32. Kotz, F., Arnold, K., Bauer, W., Schild, D., Keller, N., Sachsenheimer, K., … Rapp, B. E. [2017]. Three-dimensional printing of transparent fused silica glass. *Nature*, *544*(7650), 337–339. https://doi.org/10.1038/nature22061
33. Lubliner, J. [2008]. *Plasticity theory* (Dover ed). Mineola, N.Y: Dover Publications.
34. Luo, J., Wang, J., Bitzek, E., Huang, J. Y., Zheng, H., Tong, L., Yang, Q., Li, J., and Mao, S. X. [2016]. Size-Dependent Brittle-to-Ductile Transition in Silica Glass Nanofibers. *Nano Letters*, *16*(1), 105–113. https://doi.org/10.1021/acs.nanolett.5b03070
35. Mehta, A., and Ashish, D. K. [2020]. Silica fume and waste glass in cement concrete production: A review. *Journal of Building Engineering*, *29*, 100888. https://doi.org/10.1016/j.jobe.2019.100888
36. Muralidharan, K., Simmons, J. H., Deymier, P. A., and Runge, K. [2005]. Molecular dynamics studies of brittle fracture in vitreous silica: Review and recent progress. *Journal of Non-Crystalline Solids*, *351*(18), 1532–1542. https://doi.org/10.1016/j.jnoncrysol.2005.03.026
37. P. P., A., Nayak, D. K., Sangoju, B., Kumar, R., and Kumar, V. [2021]. Effect of nano-silica in concrete; a review. *Construction and Building Materials*, *278*, 122347. https://doi.org/10.1016/j.conbuildmat.2021.122347
24